\documentclass[12pt]{iopart}

\usepackage{iopams}
\usepackage{amssymb}
\usepackage{setstack}
\usepackage{graphicx}
\usepackage{bm}
\usepackage{color}

\begin{document}

\title{Nonequilibrium Majorana fluctuations}

\author{Sergey Smirnov}
\address{Institut f\"ur Theoretische Physik, Universit\"at Regensburg,
  D-93040 Regensburg, Germany}
\address{P. N. Lebedev Physical Institute of the Russian Academy of Sciences,
  119991 Moscow, Russia}
\ead{sergey.smirnov@physik.uni-regensburg.de}
\ead{ssmirnov@sci.lebedev.ru}

\date{\today}

\begin{abstract}
Nonequilibrium physics of random events, or fluctuations, is a unique
fingerprint of a given system. Here we demonstrate that in noninteracting
systems, whose dynamics is driven by Majorana states, the effective charge
$e^*$, characterizing the electric current fluctuations, is fractional. This
is in contrast to noninteracting Dirac systems with the trivial electronic
charge, $e^*=e$. Quite the opposite, in the Majorana state we predict two
different fractional effective charges at low and high energies, $e^*_l=e/2$
and $e^*_h=3e/2$, accessible at low and high bias voltages, respectively. We
show that while the low energy effective charge $e^*_l$ is sensitive to
thermal fluctuations of the current, the high energy effective charge $e^*_h$
is robust against thermal noise. A unique fluctuation signature of Majorana
fermions is, therefore, encoded in the high voltage tails of the electric
current noise easily accessible in experiments on strongly nonequilibrium
systems even at high temperatures.
\end{abstract}


\maketitle

\section{Introduction}
The physics of fluctuation phenomena, or noise, dating back to Brownian
\cite{Brown_1828} motion has received a systematic scientific framework after
the Einstein's \cite{Einstein_1905} and Smoluchowski's
\cite{Smoluchowski_1906} conceptual theoretical breakthrough proven
experimentally by Svedberg \cite{Svedberg_1906} and Perrin
\cite{Perrin_1908}. Spontaneous or externally excited fluctuations are an
extremely insightful tool known as the fluctuation spectroscopy. Due to their
sensitivity fluctuations scan the microscopic structure in much more detail
than mean values.

In equilibrium, nevertheless, kinetics of a given system makes a clever link
between random deviations of its physical quantities from mean values and the
mean values themselves. This link dating back to the Nyquist's
\cite{Nyquist_1928} and Callen's and Welton's \cite{Callen_1951} fundamental
discovery is known as the fluctuation-dissipation theorem \cite{Landau_V}.

In nonequilibrium the fluctuation-dissipation theorem breaks and for a given
system its fluctuation physics deviates from the mean value description. Here
nonequilibrium noise might be comparable to or, in fact, become stronger than
the equilibrium noise. It is therefore a reliable and comprehensive method to
conclusively reveal the microscopic structure of a system when measurements of
its mean quantities are physically inconclusive.

This is what currently happens in dealing with materialization of a particle
cloning its own antiparticle. Namely, via unpairing Majorana
\cite{Majorana_1937} fermions, composing a single Dirac fermion, by means of
implementations \cite{Fu_2008,Fu_2009,Lutchyn_2010,Oreg_2010} of the Kitaev's
\cite{Kitaev_2001} model it is hoped to detect a single Majorana state
\cite{Alicea_2012,Flensberg_2012,Sato_2016}. Here experiment mainly focuses on
measurements \cite{Albrecht_2016} of mean quantities such as the differential
conductance which should exhibit a peak equal to one-half of the Dirac unitary
limit \cite{Mourik_2012}. This is inconclusive because such a peak might
result, {\it e.g.}, from the Kondo effect \cite{Hewson_1997} in an asymmetric
mesoscopic system. This problem is inherent to Majorana's transport
experiments dealing with mean values. Nevertheless, it is possible to get a
conclusive signature of Majorana fermions from the mean value description of
both Majorana transport
\cite{Fidkowski_2012,Kundu_2013,Ilan_2014,Lobos_2015,Wang_2016,Lutchyn_2017}
and Majorana thermodynamics \cite{Smirnov_2015}.

The freedom to involve nonequilibrium noise
\cite{Liu_2015,Liu_2015a,Beenakker_2015,Valentini_2016} in the fluctuation
spectroscopy of Majorana fermions triggers transport experiments on Majorana
physics to a new azimuth and makes it more interesting. This is because, as
mentioned above, fluctuations are usually conclusive on the microscopic
structure of a system and at the same time these are transport experiments
which are in general simpler than thermodynamic ones. So far Majorana noise
has mainly been discussed in linear response. However, the real beauty of
nonequilibrium noise is still to be explored beyond linear response. Here
fluctuations of the electric current may be characterized by the so-called
effective charge $e^*$ which is not directly related to a particle's
elementary charge but rather characterizes backscattering processes
\cite{Sela_2006}. Modern experiments \cite{Ferrier_2016} have already reached
a remarkable accuracy and enabled one to measure the noise of the electric
current providing $e^*$ as a unique fluctuation fingerprint of the system.

In the present work we explore strongly nonequilibrium fluctuations of the
electric current flowing through a noninteracting quantum dot coupled to a
topological superconductor supporting at its ends two Majorana bound states
implemented via the Kitaev's chain model. It is well known that in the absence
of Majorana fermions the effective charge for a noninteracting quantum dot is
trivial and identical to the electronic charge, $e^*=e$. Here we demonstrate
that in the presence of Majorana fermions 1) the effective charge
fractionalizes to 2) $e^*_l=e/2$ at low energies, to 3) $e^*_h=3e/2$ at high
energies and show that 4) even when the low energy effective charge
$e^*_l=e/2$ is washed out by, {\it e.g.}, thermal noise, the high energy
effective charge $e^*_h=3e/2$ is robust and persists up to very high
temperatures providing a simple and reliable experimental platform for a
unique signature of Majorana fermions out of strongly nonequilibrium
fluctuations.

The paper is organized as follows. In Section \ref{Thr_setup} we present a
Majorana setup suitable for experiments on nonequilibrium noise and explore it
using the Keldysh field integral framework. The results on nonequilibrium
noise, in particular, on the effective charge are shown and discussed in
Section \ref{RD}. We conclude with Section \ref{Conclusion}. \ref{appA} and
\ref{appB} provide details on the Keldysh field integral in the presence of
Majorana fermions.
\section{Theoretical setup and its Keldysh field integral description}\label{Thr_setup}
Let us consider a setup similar to the one of Ref. \cite{Smirnov_2015}. It
represents a noninteracting quantum dot coupled via tunneling interaction
to two ($L$ and $R$) noninteracting contacts. In contrast to the equilibrium
setup of Ref. \cite{Smirnov_2015}, here the contacts may be used to apply a
\begin{figure}
\centering
\includegraphics[width=10.0 cm]{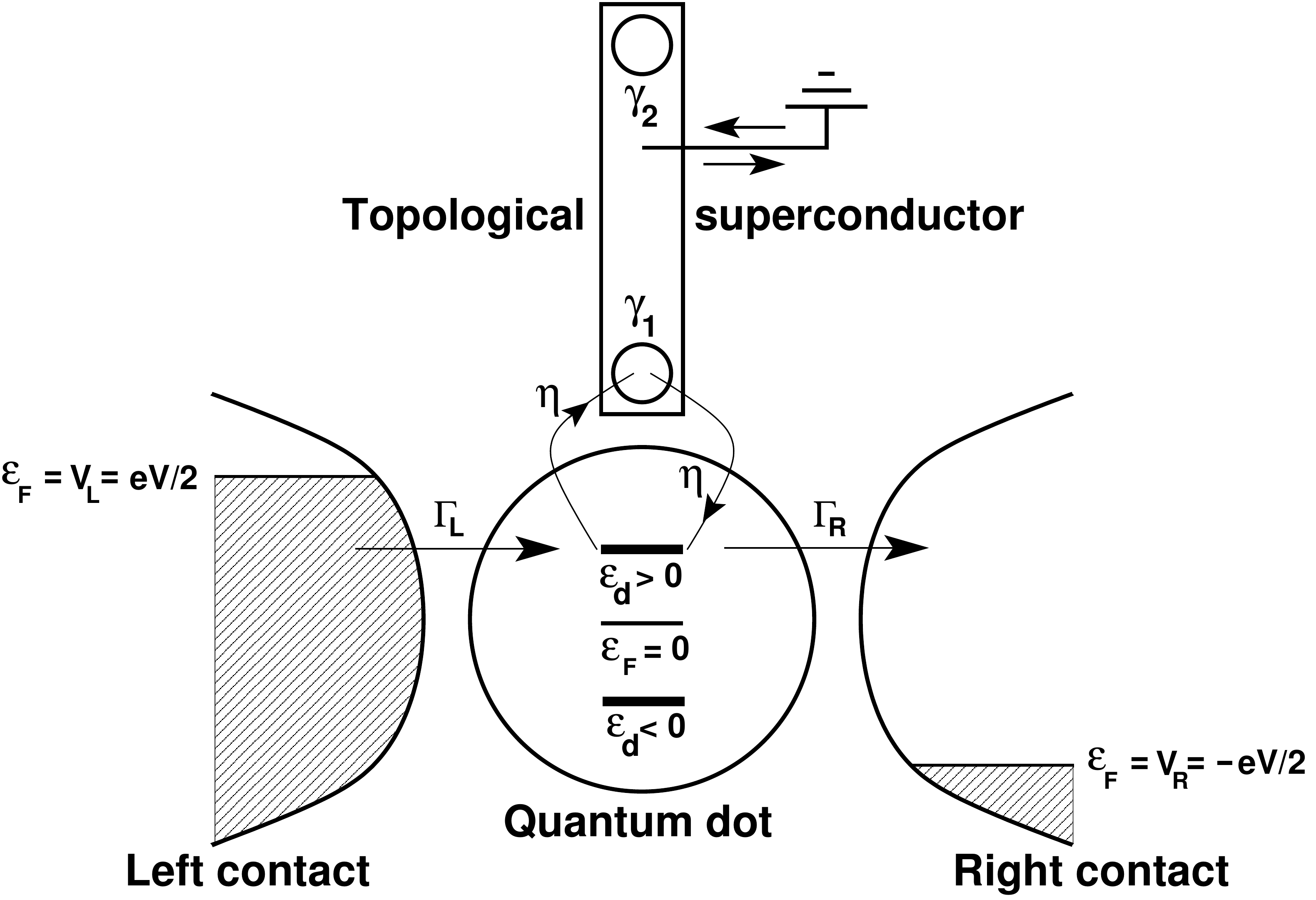}
\caption{\label{figure_1} Quantum dot with a single particle nondegenerate
  energy level $\epsilon_d$ is linked via tunneling mechanisms to two normal
  contacts and to one end of a one-dimensional topological  superconductor
  supporting two Majorana bound states, $\gamma_1$, $\gamma_2$, at its
  ends. Here $\Gamma_L$, $\Gamma_R$ characterize the tunneling strength
  between the left (L) and right (R) contacts while $\eta$ is the strength of
  the tunneling between the quantum dot and the Majorana bound state
  $\gamma_1$. A bias voltage $V$ may be applied to the contacts and induce an
  electric current flowing in the direction of arrows. The electric current
  $<I(t)>$ and its noise $<I(t)I(t')>$ may be measured in one of the contacts,
  {\it e.g.}, in the left contact, $<I_L(t)>$, $<I_L(t)I_L(t')>$.}
\end{figure}
bias voltage $V$ to the quantum dot, $V_L=-V_R=V/2$, $V_L-V_R=V$, as it is
schematically shown in Fig. \ref{figure_1} for the case $V<0$. The quantum dot
has a one single particle level which is spin nondegenerate as may be
experimentally implemented, {\it e.g.}, via the Zeeman splitting which also
filters out the Kondo effect \cite{Hewson_1997,Smirnov_2013}. Note also that
below we explore strongly nonequilibrium states which are accessed at very
high bias voltage $V$ so that the Kondo state is totally ruined
\cite{Smirnov_2013a} in any case and, therefore, does not lead to any
experimental ambiguity. Finally, similar to Ref. \cite{Smirnov_2015}, the
quantum dot interacts via another tunneling mechanism with a grounded
topological superconductor supporting at its ends two Majorana bound states.

To give the problem a concrete and mathematically convenient treatment we
formulate it in terms of the quantum many-particle Keldysh Lagrangian
$\mathcal{L}_\mathcal{K}(q,p)=\sum_ii\hbar q_i\dot{p}_i-\mathcal{H}_\mathcal{K}(q,p)$ 
which constitutes the basis for the Keldysh action,
$\mathcal{S}_\mathcal{K}$. Here the momenta $p_i$ and coordinates $q_i$ are
the fermionic coherent states of the system and their conjugate partners,
respectively. The Lagrangian formulation is fully equivalent to the quantum
Hamiltonian $\mathcal{H}_\mathcal{K}(q,p)$ formulation but has a certain
technical advantage in calculating strongly nonequilibrium fluctuations of the
electric current via the Keldysh field integral \cite{Altland_2010} which we
employ below to obtain the current-current correlation function.

For the quantum dot ($p=\psi$, $q=\bar{\psi}$), contacts ($p=\phi$,
$q=\bar{\phi}$) and tunneling between them we, respectively, have:
\begin{equation}
\mathcal{H}_{\rm QD}(q,p)=\epsilon_d\bar{\psi}(t)\psi(t),
\label{H_QD}
\end{equation}
where $\epsilon_d$ is the quantum dot energy level and the real time $t$ runs
along the Keldysh closed contour $\mathcal{C}_\mathcal{K}$,
$t\in\mathcal{C}_\mathcal{K}$,
\begin{equation}
\mathcal{H}_{\rm C}(q,p)=\sum_{l=\{L,R\};k_l}\epsilon_{lk_l}\bar{\phi}_{lk_l}(t)\phi_{lk_l}(t),
\label{H_C}
\end{equation}
where below we will assume identical quantum numbers $k_l$ in both $L$ and $R$
contacts as well as large contacts so that their spectrum $\epsilon_{lk_l}$ is
continuous and their density of states $\nu_{\rm C}$ is constant in the
vicinity of the Fermi energy,
\begin{equation}
\mathcal{H}_{\rm DT}(q,p)=\sum_{l=\{L,R\};k_l}T_{lk_l}\bar{\phi}_{lk_l}(t)\psi(t)+{\rm H.c.}
\label{H_DT}
\end{equation}
We use for the Dirac tunneling the standard assumption that the tunneling
matrix elements weakly depend on the contacts quantum numbers,
$T_{lk_l}\approx T_l$. This allows to characterize the tunneling coupling
by the energy scales $\Gamma_l=\pi\nu_{\rm C}|T_l|^2$ or
$\Gamma\equiv\Gamma_L+\Gamma_R$.

Finally, the topological ($p=\zeta$, $q=\bar{\zeta}$) part of the Hamiltonian
is given as the sum of the Hamiltonians of the topological superconductor and
its tunneling interaction with the quantum dot: 
\begin{equation}
\mathcal{H}_{\rm TS}(q,p)=i\xi\bar{\zeta}_2(t)\zeta_1(t)/2,
\label{H_TS}
\end{equation}
where $\xi$ is the energy originating from the overlap of the Majorana bound
states $\zeta_1$ and $\zeta_2$ (in particular, $\xi=0$ if there is no overlap
as in sufficiently long Kitaev's chains),
\begin{equation}
\mathcal{H}_{\rm MT}(q,p)=\eta^*\bar{\psi}(t)\zeta_1(t)+{\rm H.c.},
\label{H_MT}
\end{equation}
where the Majorana tunneling entangles the Dirac fermions of the quantum dot
with only one Majorana state, $\zeta_1$, and is characterized by the energy
scale $|\eta|$.

Due to the fundamental property of the Majorana fields,
$\bar{\zeta}_j(t)=\zeta_j(t)$, $j=1,2$, and the canonic fermionic
anticommutation relations the field integral for the Keldysh partition
function,
\begin{equation}
\mathcal{Z}_\mathcal{K}=\int\mathcal{D}(p,q)\exp\biggl(\frac{i}{\hbar}\mathcal{S}_\mathcal{K}\biggl),
\label{Z_K}
\end{equation}
is a functional integral with the constraints
$\bar{\zeta}_j(t)=\zeta_j(t)$, $\zeta^2_j(t)=1$ which might be viewed as
fermionic constraints imposed at any given discrete time of the Keldysh closed
contour. Nevertheless, since $\mathcal{H}_\mathcal{K}$ is quadratic in all
$p_i$, $q_i$, this field integral may be solved exactly as in many standard
textbooks \cite{Altland_2010} as explained in detail in \ref{appA} and
\ref{appB}.

Using an imaginary time field theory, it has been rigorously proven
\cite{Smirnov_2015} by entropic reasoning that a macroscopic state of the
above setup is Majorana dominated at low temperatures. Therefore, it must
exhibit various fractionalizations \cite{Sato_2016} of its observables. Here,
in particular, we are interested in fractionalizations of the electric current
fluctuations.

To this end we introduce suitable sources into the Keldysh partition function
(\ref{Z_K}) turning it into the Keldysh generating functional:
\begin{equation}
\mathcal{J}_\mathcal{K}=\int\mathcal{D}(p,q)\exp\biggl(\frac{i}{\hbar}\mathcal{S}_\mathcal{K}
+\sum_{l=\{L,R\}}\int_{\mathcal{C}_\mathcal{K}}\!\!\!\!\!dt\,J_l(t)I_l(t)\biggl),
\label{J_L}
\end{equation}
where $J_l(t)$ is the source field and $I_l(t)$ is the electric current
field. The mean current and current-current correlator are then obtained by
proper functional differentiations of Eq. (\ref{J_L}) with respect to the
source field. Here, calculating the current-current correlator, one should
remember that due to the topological superconductor (or fermionic constraints)
various anomalous expectation values do not vanish, {\it i.e.}, in general
$\langle\psi_1\psi_2\bar{\psi}_3\bar{\psi}_4\rangle$ also includes the term
$\langle\psi_1\psi_2\rangle\langle\bar{\psi_3}\bar{\psi_4}\rangle$.

Below we are interested in the so called greater current-current correlator
$S^>(t,t')\equiv\langle \delta I_L(t)\delta I_L(t')\rangle$ (as opposed to the
lesser one, $S^<(t,t')\equiv\langle \delta I_L(t')\delta I_L(t)\rangle$),
where $\delta I_L(t)\equiv I_L(t)-I_L(V)$ is the electric current fluctuation
field and $I_L(V)\equiv\langle I_L(t)\rangle$ is the mean electric
current. More precisely, we calculate the Fourier component $S^>(\omega,V)$ as
a function of the bias voltage $V$ at zero frequency,
$S^>(V)\equiv S^>(\omega=0,V)$. 

The essential and experimentally relevant characteristic of the electric
current fluctuations is the effective charge $e^*$ which relates the nonlinear
parts of $S^>(V)$ and $I(V)\equiv I_L(V)$. In modern experiments
\cite{Ferrier_2016} on nonequilibrium current fluctuations in quantum dots one
measures the shot noise, $S^>(V)$, and the mean current, $I(V)$, subtracts the
linear parts to get the corresponding nonlinear quantities $S^>_K(V)$,
$I_K(V)$ and finally obtains the ratio $e^*/e=S^>_K(V)/e|I_K(V)|$ at small
bias voltages. In particular, for noninteracting quantum dots one gets the
trivial result, $e^*=e$.

We generalize the above definition of the effective charge in such a way that
it reproduces not only the standard definition at small bias voltages but also
provides a unique fluctuation fingerprint of a system far from equilibrium
where the system's dynamics is highly nonlinear and the expansion in powers of
$V$ makes no sense at all. To this end we note that at small values of $|V|$
the nonlinear parts of both the shot noise and the mean current are cubic in
$V$. Therefore, at low voltages the second derivatives $d^2S^>(V)/dV^2$ and
$d^2I(V)/dV^2$ are linear in $V$ and thus linearly depend on each other with
the ratio $[d^2S^>(V)/dV^2]/|d^2[eI(V)]/dV^2|=S^>_K(V)/e|I_K(V)|$. Therefore,
we define the effective charge as:
\begin{equation}
\frac{e^*}{e}=\frac{\frac{d^2S^>(V)}{dV^2}}{\big|\frac{d^2[eI(V)]}{dV^2}\big|},
\label{Eff_Ch}
\end{equation}
which is applicable when the second derivatives linearly depend on each
other. Note, that the linear dependence of $d^2S^>(V)/dV^2$ on $d^2I(V)/dV^2$
does not necessarily imply a linear dependence of these derivatives on the
bias voltage. In fact, at large bias voltages expansions in powers of $V$ do
not exist while $d^2S^>(V)/dV^2$ may still linearly depend on $d^2I(V)/dV^2$
and, therefore, the effective charge in Eq. (\ref{Eff_Ch}) makes sense even at
extremely high bias voltages.
\begin{figure}
\centering
\includegraphics[width=10.0 cm]{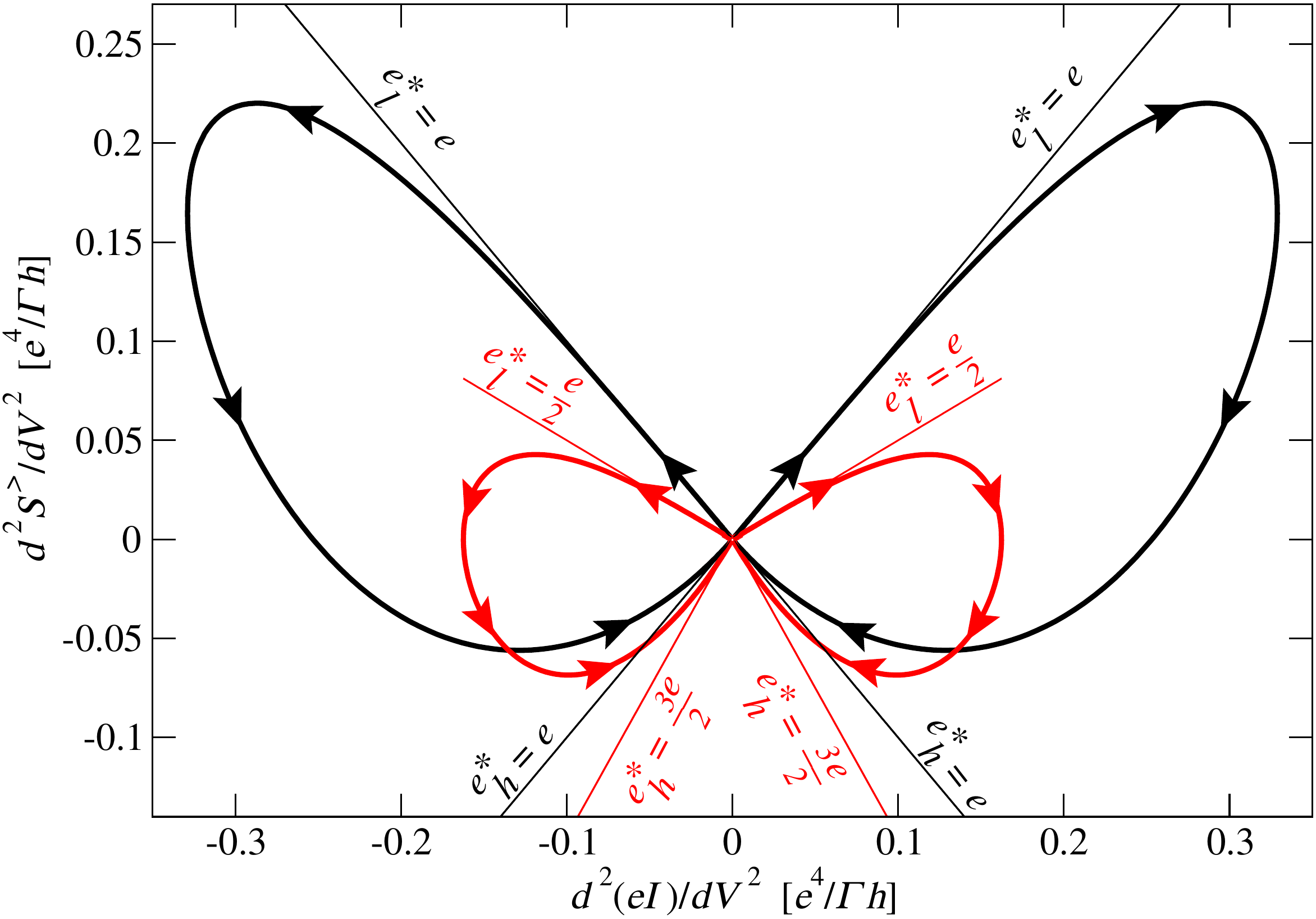}
\caption{\label{figure_2} The second derivative of the current noise,
  $d^2S^>(V)/dV^2$, as a function of the second derivative of the mean current
  times the electronic charge $e$, $d^2[eI(V)]/dV^2$. The curves are
  parameterized by the bias voltage $V$ which grows in the direction of the
  arrows from $e|V|/\Gamma=10^{-3}$ to $e|V|/\Gamma=10$. For both curves the
  temperature is the same, $k_{\rm B}T/\Gamma=10^{-6}$. The black curve is
  for $\xi/\Gamma=10^2$. The red curve is for $\xi/\Gamma=10^{-4}$.}
\end{figure}

Note also that the definition in Eq. (\ref{Eff_Ch}) is highly consistent
because, as shown below, in the absence of Majorana fermions it gives for the
noninteracting case $e^*=e$ both at low and very large bias
voltages. Importantly, this definition is also highly relevant for experiments
because each of the second derivatives may be measured with sufficient
accuracy already at present.
\section{Results and discussion}\label{RD}
Let us consider the situation when $|\eta|>\Gamma$. We also currently assume
$\epsilon_d=0$ and $\Gamma_L=\Gamma_R$. Fig. \ref{figure_2} shows
$d^2S^>(V)/dV^2$ as a function of $d^2I(V)/dV^2$ when $|\eta|=8\Gamma$ for two
different values of the overlap energy $\xi$. The black curve is for
$\xi/\Gamma=10^2$. In this case the Majorana fermions strongly overlap forming
a single Dirac fermion leading to the current fluctuations with a trivial
effective charge equal to the electronic charge both at low ($e|V|\ll\Gamma$)
and high ($e|V|>\Gamma$) energies. Indeed, the curve is linear near the origin
both at its starting point and at its ending point with the tangent lines
having unit absolute slope resulting in $e^*_l=e^*_h=e$. However, when
$\xi/\Gamma=10^{-4}$, Majorana bound states overlap weakly and the fluctuation
\begin{figure}
\centering
\includegraphics[width=10.0 cm]{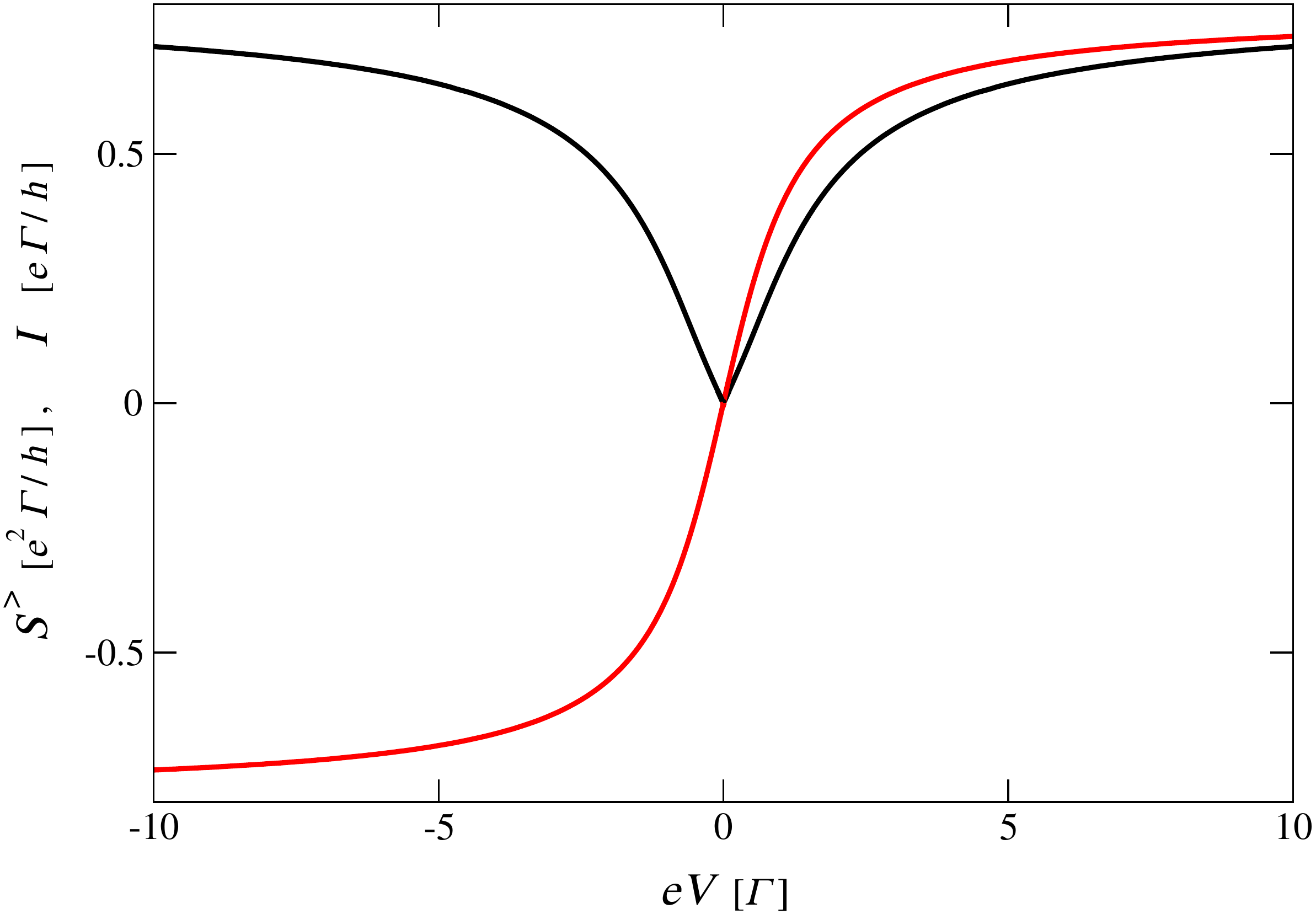}
\caption{\label{figure_3} The current noise $S^>(V)$ (black, symmetric) and
  the mean current $I(V)$ (red, antisymmetric) as functions of the bias
  voltage $V$. For both curves $k_{\rm B}T/\Gamma=10^{-6}$ and
  $\xi/\Gamma=10^{-4}$. For very small voltages $k_BT\ll e|V|\ll\Gamma$ both
  curves are almost linear with the slopes $dS^>(V)/dV=0.25$, $dI(V)/dV=0.5$
  in full accordance with Refs. \cite{Liu_2015,Liu_2015a} (Note that here we
  calculate the greater noise and not the symmetrized noise as in
  Refs. \cite{Liu_2015,Liu_2015a}. For zero frequency they differ by a factor
  of 2).}
\end{figure}
physics is governed by fractional degrees of freedom leading to fractional
effective charges at low and high energies. In this case the curve is linear
near the origin both at its starting point and at its ending point with the
tangent lines having, respectively, absolute slopes equal to 1/2 and 3/2
resulting in $e^*_l=e/2$ at low energies ($e|V|\ll\Gamma$) and $e^*_h=3e/2$ at
high energies ($|\eta|>e|V|>\Gamma$). At voltages $e|V|\gg|\eta|$ the Majorana
state is ineffective and the curve acquires the trivial linear character with
$e^*=e$ which is not visible in Fig. \ref{figure_2} because both of the second
derivatives become very small at the high voltage tails of $S^>(V)$ and $I(V)$
shown in Fig. \ref{figure_3} for the case $\xi/\Gamma=10^{-4}$. However, it
becomes visible when the effective charge is plotted as a function of $V$ (see
Fig. \ref{figure_5} below).
\begin{figure}
\centering
\includegraphics[width=10.0 cm]{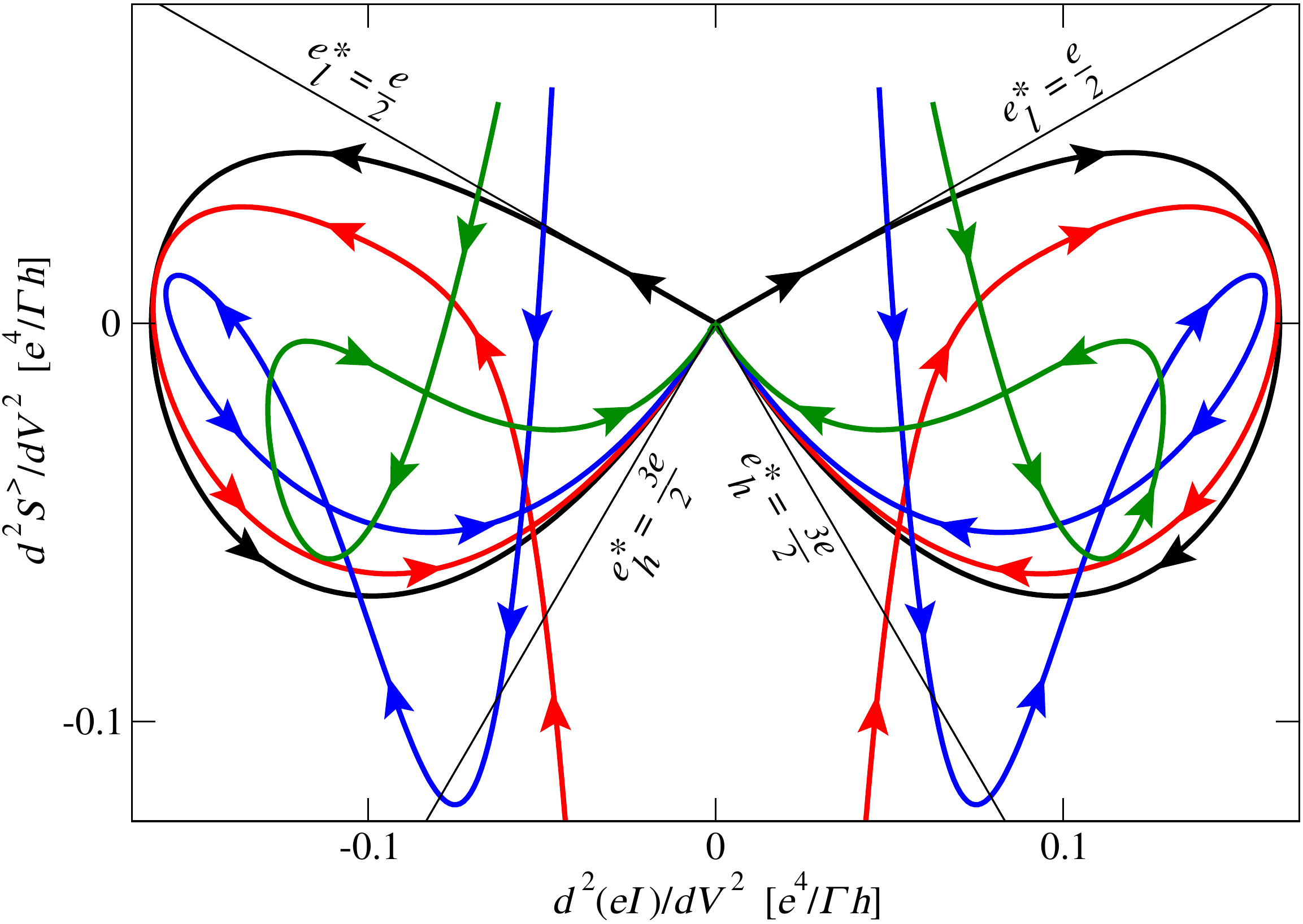}
\caption{\label{figure_4} The second derivative of the current noise,
  $d^2S^>(V)/dV^2$, as a function of the second derivative of the mean current
  times the electronic charge $e$, $d^2[eI(V)]/dV^2$. For all the curves
  $\xi/\Gamma=10^{-4}$. The temperatures are $k_{\rm B}T/\Gamma=10^{-6}$
  (black), $k_{\rm B}T/\Gamma=10^{-2}$ (red),
  $k_{\rm B}T/\Gamma=3\cdot10^{-2}$ (blue) and $k_{\rm B}T/\Gamma=10^{-1}$
  (green). The curves are parameterized by the bias voltage $V$ which grows in
  the direction of the arrows from $e|V|/\Gamma=10^{-3}$ to $e|V|/\Gamma=10$
  (black), from $e|V|/\Gamma=8.85\cdot10^{-2}$ to $e|V|/\Gamma=10$ (red), from
  $e|V|/\Gamma=1.025\cdot10^{-1}$ to $e|V|/\Gamma=10$ (blue) and from
  $e|V|/\Gamma=2.15\cdot10^{-1}$ to $e|V|/\Gamma=10$ (green). All the high
  temperature curves start from positive values of $d^2S^>(V)/dV^2$. For the
  red curve the starting point is chosen so as not to overload the figure with
  its low voltage or thermal noise branch going from large positive to large
  negative values of $d^2S^>(V)/dV^2$ since in the present research we do not
  focus on the thermal Majorana noise.}
\end{figure}

In Fig. \ref{figure_4} we show $d^2S^>(V)/dV^2$ as a function of
$d^2I(V)/dV^2$ for $|\eta|=8\Gamma$, $\xi/\Gamma=10^{-4}$ for different
temperatures. Since the overlap energy is small, the current fluctuations are
essentially governed by the Majorana degrees of freedom. Here we increase the
bias voltage of the starting points of the high temperature curves to stay in
the regime $e|V|>k_{\rm B}T$ in order to avoid high values of the thermal
Majorana noise which is not in the focus of the present research. At high
temperatures (red, blue and green curves) the low energy effective charge
$e^*_l=e/2$ is completely washed out by thermal fluctuations of the electric
current. However, the high energy effective charge $e^*_h=3e/2$ is robust
against thermal noise and persists up to very high temperatures,
$k_{\rm B}T/\Gamma=10^{-1}$ (green curve).

Let us estimate the temperature at which the fractional high energy effective
charge $e^*_h=3e/2$ might be observed in experiments. If the induced
superconducting gap is taken from Ref. \cite{Mourik_2012},
$\Delta=250\,\mu{\rm eV}$, and $|\eta|\approx\Delta$, then we obtain
$T\approx 36\,{\rm mK}$ which is easily reachable in modern experiments. If
the induced superconducting gap is taken from Ref. \cite{Wang_2013},
$\Delta=15\,{\rm meV}$, then $T\approx 2\,{\rm K}$ which is even more
reachable.

The fractional high energy effective charge $e^*_h=3e/2$ is perfectly achieved
only at $|\eta|>e|V|\gg\Gamma$ which requires $|\eta|\gg\Gamma$. However,
according to our numerical analysis we estimate that for $|\eta|>\Gamma$ (and
small $\xi$) it weakly deviates from the value $3e/2$. Namely, from numerical
fitting we get $e^*_h\approx [3/2-2(\Gamma/\eta)^2]e$. So that
$|\eta|=8\Gamma$ gives $e^*_h\approx 1.47e$, $|\eta|=20\Gamma$ gives
$e^*_h\approx 1.495e$ and $|\eta|=50\Gamma$ gives $e^*_h\approx 1.4992e$.

Importantly, by means of a gate voltage one may easily in realistic
experiments increase $\epsilon_d$ so that $\epsilon_d>0$,
$|\epsilon_d|>\Gamma$. In this case the quantum dot is in the empty orbital
regime \cite{Hewson_1997} opposite to the Kondo one. In this way one fully
eliminates \cite{Vernek_2014} the Kondo effect. At the same time $e_l^*$ and
$e_h^*$ do not change as soon as the quantum dot is in the Majorana universal
regime, $|\eta|>{\rm max}\{|\epsilon_d|, \Gamma, e|V|\}$. Since $\Gamma$
and/or $|\eta|$ may be easily varied in modern experiments
\cite{Goldhaber-Gordon_1998}, the Majorana universal regime is readily
reachable in modern laboratories. Therefore, one may unambiguously observe in
realistic experiments the universal plateaus $e_l^*=e/2$ and $e_h^*=3e/2$ in
the empty orbital and Majorana universal regime as it is shown in
Fig. \ref{figure_5} for the case $|\eta|/\Gamma=10^3$ and
$\epsilon_d/\Gamma=8$. These plateaus are universal and do not depend on
$\epsilon_d$ as soon as $|\epsilon_d|<|\eta|$. Also for $|\eta|/\Gamma=10^2$
the plot on Fig. \ref{figure_5} is almost unchanged. Moreover, we find that
the $e^*_h$ plateau survives up to very high temperatures,
$k_BT\sim10^{-2}|\eta|$, {\it i.e.}, up to $k_BT=10\,\Gamma$ for the present
case as shown by the black dashed line in Fig. \ref{figure_5}. As one can see,
although the plateau $e^*_h$ becomes very narrow at such a high temperature,
it is still visible and it almost reaches the value $3e/2$ even at
$k_BT=10\,\Gamma$.
\begin{figure}
\includegraphics[width=10.0 cm]{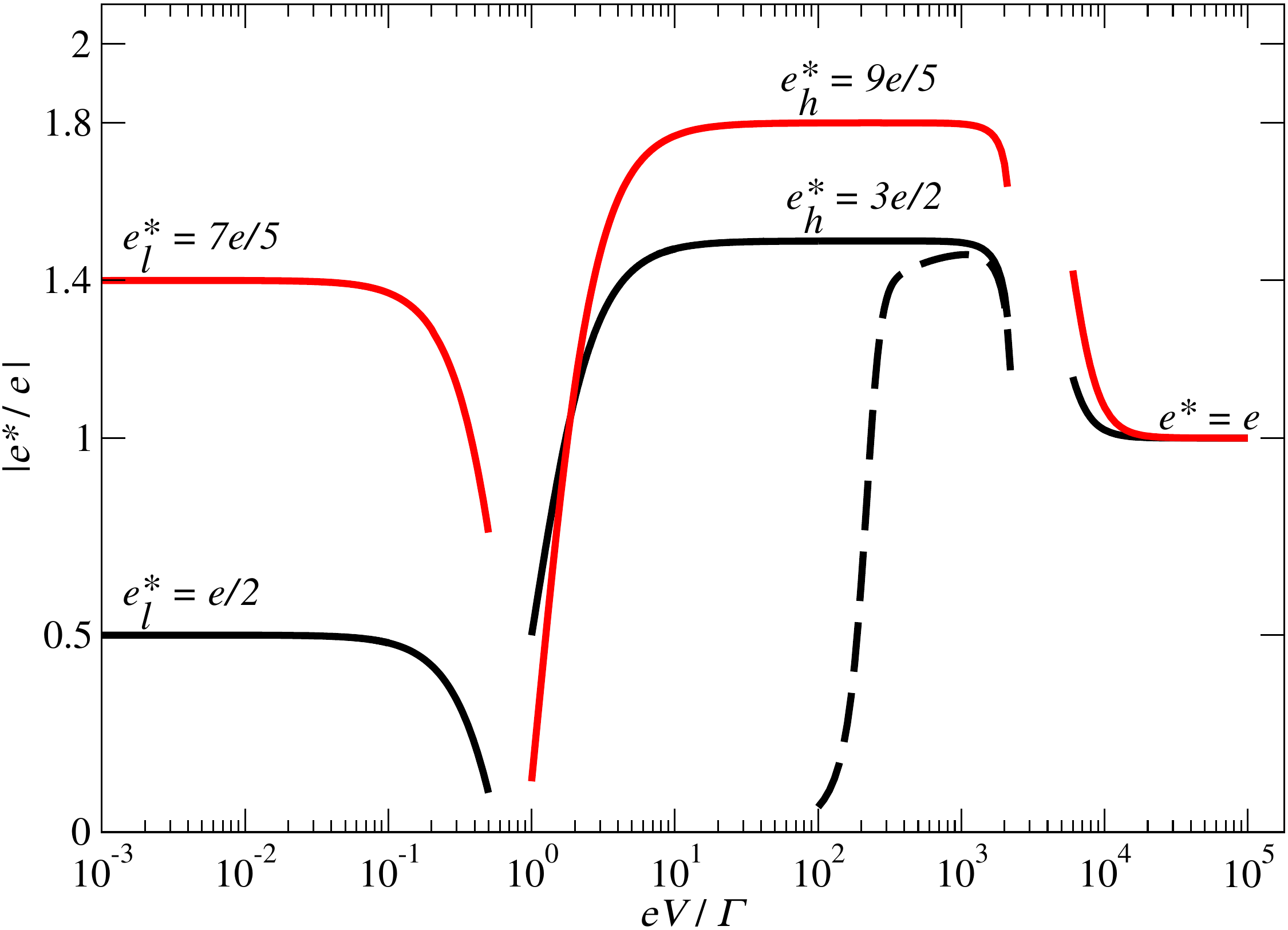}
\centering
\caption{\label{figure_5} Effective charge as a function of the bias
  voltage. The black solid line corresponds to the case
  $\gamma_L=\gamma_R=0.5$ while the red solid line corresponds to the case
  $\gamma_L=0.8$, $\gamma_R=0.2$. For both solid curves
  $k_BT/\Gamma=10^{-6}$. The black dashed line corresponds to the case
  $\gamma_L=\gamma_R=0.5$, $k_BT/\Gamma=10^{-2}|\eta|=10\,\Gamma$. In all cases
  the Majorana bound states overlap weakly, $\xi/\Gamma=10^{-4}$, {\it i.e.},
  they are well defined and govern the nonequilibrium physics up to very high
  voltages $e|V|\sim|\eta|$. For extremely high voltages, $e|V|\gg|\eta|$, the
  Majorana state is ineffective. As a result, at $e|V|\gg|\eta|$ the effective
  charge becomes equal to the trivial electronic charge, $e^*=e$.}
\end{figure}

Another important aspect is the universality of the effective charge plateaus
$e^*_l$ and $e^*_h$ when the quantum dot is asymmetrically coupled to the left
and right contacts. This asymmetry may be characterized by the quantities
$\gamma_L\equiv\Gamma_L/\Gamma$, $\gamma_R\equiv\Gamma_R/\Gamma$, which
satisfy $\gamma_L+\gamma_R=1$. The symmetric setup discussed above corresponds
to the case $\gamma_L=\gamma_R=0.5$. In a general setup
$\gamma_L\neq\gamma_R$. Nevertheless, the effective charge $e^*$ in
Eq. (\ref{Eff_Ch}) is characterized by two different universal plateaus
$e^*_l$ and $e^*_h$ at low and high bias voltages, respectively. In this
general asymmetric situation when $\gamma_L\neq\gamma_R$ and when $\xi$ is
small, {\it i.e.}, the two Majorana bound states are well separated, the low
energy and high energy plateaus of the effective charge are:
\begin{equation}
e^*_l=(3\gamma_L-1)e,\quad e^*_h=(1+\gamma_L)e.
\label{eleh}
\end{equation}
We obtain these values with any desired numerical precision which means that
Eq. (\ref{eleh}) is the numerically exact result. Its analytical proof is a
complicated task especially in the case of $e^*_h$ taking place at high
voltages where the dynamics is nonlinear. This analytical proof could be based
on a semiclassical picture \cite{Haim_2015} and will be a challenge for our
future research which should, in particular, explain the physical meaning of
the high energy effective charge $e^*_h$ predicted currently by different
numerical techniques with very high precision.

From Eq. (\ref{eleh}) one can see that $e^*_h-e^*_l=2(1-\gamma_L)$. Only when
$\gamma_L\rightarrow 1$, one gets a unique effective charge both at low and
high energies, $e^*_l=e^*_h=2e$. However, as soon as $0\leqslant\gamma_L<1$,
the unique value, $2e$, of the effective charge splits into two different
values. In the symmetric case, $\gamma_L=\gamma_R=0.5$, one gets from
Eq. (\ref{eleh}) the result discussed above, $e^*_l=e/2$,
$e^*_h=3e/2$. However, when, for example, $\gamma_L=0.8$, $\gamma_R=0.2$, one
gets from Eq. (\ref{eleh}) $e^*_l=7e/5$, $e^*_h=9e/5$, as shown in
Fig. \ref{figure_5}. Once again, we emphasize that the low energy, $e^*_l$,
and high energy, $e^*_h$, effective charge plateaus, given by
Eq. (\ref{eleh}), are universal for all possible values of the asymmetries
$\gamma_L$, $\gamma_R$. In particular they do not depend on $\epsilon_d$,
{\it i.e.}, they do not depend on the gate voltage.

On the other side, when $\xi$ is large, {\it i.e.}, when the two Majorana
bound states strongly overlap forming a single Dirac fermion, we obtain with
any desired numerical precision that $e^*_l=e^*_h=e$ for all $\gamma_L$,
$\gamma_R$ except for a small vicinity of the point $\gamma_L=1$,
$\gamma_R=0$, where $e^*_l$ and $e^*_h$ are sharply peaked to the value
$e^*_l=e^*_h=2e$ reached exactly at the point $\gamma_L=1$,
$\gamma_R=0$. Therefore, in the case of large $\xi$, when the Majorana
fermions form a single Dirac fermion, the effective charge plateaus, given by
Eq. (\ref{eleh}), do not appear for any degree of asymmetry described by the
values of $\gamma_L$ and $\gamma_R$.

This shows that the presence of two different universal effective charges at
low and high energies, $e^*_l$ and $e^*_h$, respectively, whose values are
given by Eq. (\ref{eleh}), is a unique fluctuation signature of the presence
of Majorana fermions in the topological superconductor independent of the
asymmetry in the coupling of the quantum dot to the left and right
contacts. An experimental detection of at least one of these effective charges
is enough to conclusively claim that the topological superconductor in this
setup supports Majorana fermions.

Note also one practical aspect of Eq. (\ref{eleh}). As soon as one of the
effective charges, $e^*_l$ or $e^*_h$, is detected in an experiment, the
asymmetries $\gamma_L$ and $\gamma_R$ immediately follow in a simple way from
Eq. (\ref{eleh}). This simple way of extraction of $\gamma_L$ and $\gamma_R$
is definitely a practical advantage since usually in experiments it is
difficult to measure the values of $\gamma_L$ and $\gamma_R$. At the same time
it is often necessary to know the values of $\gamma_L$ and $\gamma_R$ to
theoretically describe realistic experiments.
\begin{figure}
\includegraphics[width=10.0 cm]{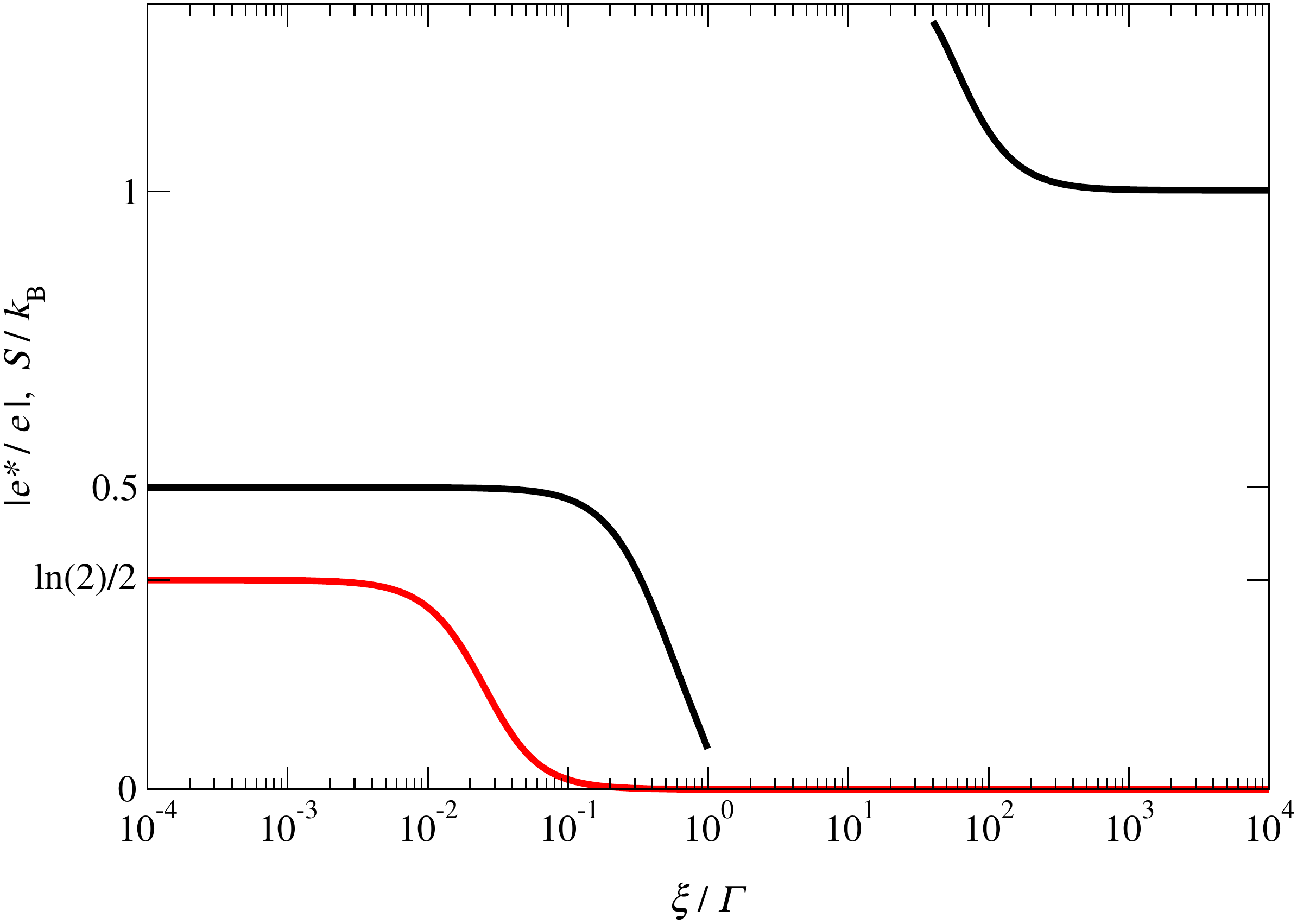}
\centering
\caption{\label{figure_6} The effective charge $|e^*/e|$ (black line) as a
  function of the overlap energy $\xi$ for $\epsilon_d/\Gamma=8$,
  $k_BT/\Gamma=10^{-8}$, $|\eta|/\Gamma=10^3$, $e|V|/\Gamma=10^{-2}$. The red
  line is the system's entropy $S/k_B$ as a function of the overlap energy
  $\xi$ for the same parameters but with zero bias voltage, $V=0$. When the
  entropy is equal to the half-fermionic value, $S/k_B=\ln(2)/2$, the Majorana
  modes are well separated and they govern the equilibrium macroscopic state
  of the system. Weak deviations from this Majorana equilibrium macroscopic
  state occur at small bias voltages $k_BT\ll e|V|\ll\Gamma$. At these bias
  voltages the low energy effective charge is equal to one-half of the
  electronic charge, $e^*_l=e/2$. At very large values of $\xi$ the two
  Majorana modes strongly overlap and form one Dirac fermion. As a result, the
  entropy drops to the zero value corresponding to one nondegenerate ground
  state, namely the state with the quantum dot having zero electrons
  ($\epsilon_d>0$). Weak deviations from this trivial equilibrium macroscopic
  state are characterized by the trivial value of the effective charge,
  $e^*=e$, when the bias voltage is small, $k_BT\ll e|V|\ll\Gamma$.}
\end{figure}

We would like to emphasize that the low energy effective charge $e^*_l$ is
obtained from Eq. (\ref{Eff_Ch}) at $k_BT\ll e|V|\ll\Gamma$. Although the
voltage is small here, $e|V|\ll\Gamma$, it is still finite to make thermal
noise insignificant, $e|V|\gg k_BT$. Therefore, the system is not in
equilibrium. To understand how far it is from the equilibrium and to which
extent its equilibrium macroscopic states may still govern the behavior of
$e^*_l$ we compare the behavior of $e^*_l$ at $k_BT\ll e|V|\ll\Gamma$ with the
behavior of the system's entropy at $V=0$. Here it has been rigorously proven
\cite{Smirnov_2015} that the macroscopic state of the present setup is
characterized by the entropy plateau $S=\ln(2)/2$. This shows that the
macroscopic state consists of non-integer number of microscopic states namely
it consists of one-half of the Dirac fermion state. That the Majorana
equilibrium macroscopic state indeed governs the behavior of $e^*_l$ is clear
from the following fact. When $\xi$ grows, the two Majorana fermions combine
into a single Dirac fermion and Eq. (\ref{Eff_Ch}) gives a transition from the
plateau $e^*_l=e/2$ to the plateau with the integer electronic charge
$e^*=e$. At the same time, when $\xi$ grows, the Majorana plateau $S=\ln(2)/2$
is fully ruined to the trivial plateau $S=0$ as shown in Fig. \ref{figure_6}.

Concerning the high energy effective charge we would like to note that its
presence is also a unique signature of the Majorana fermions for all
$0<\gamma_L<1$. This is particularly clear in the case
$\gamma_L=\gamma_R=0.5$. Here the noise properties characterized by
$e_h^*=3e/2$ cannot be induced by two particle processes as one would expect
from the standard point of view where an effective charge is usually
associated with backscattering processes at $V\rightarrow 0$. From this
traditional perspective one would conclude that $e^*_h=3e/2$ is the result of
a combination of single particle and two particle processes due to Andreev
reflection. However, this traditional approach is usually applied at
$V\rightarrow 0$ \cite{Sela_2006,Ferrier_2016} and its adequacy at large bias
voltages, where the dynamics is highly nonlinear, would be a question for
future research, especially in connection with the definition of the effective
charge given by Eq. (\ref{Eff_Ch}). In the present case, however, this
traditional point of view is definitely inapplicable because for
$\gamma_L=\gamma_R=0.5$ the Andreev current, the only possible source of two
particle processes here, is equal to zero \cite{Liu_2015a}. This shows that
the traditional explanation of the high energy effective charge in terms of
combination of different processes is meaningless and the value $e_h^*=3e/2$
is of pure Majorana nature. Likewise, when $\gamma_L\neq\gamma_R$, the
effective charge $e^*_h$ is also of Majorana nature although the Andreev
current may be finite in this situation. Indeed, when $\gamma_L\neq\gamma_R$
the Majorana nature of the high energy effective charge is obvious from the
fact that it is fractional for small values of $\xi$, when the Majorana modes
are well separated, but takes the trivial value $e^*=e$ as soon as the two
Majorana fermions combine into a single Dirac fermion at large values of
$\xi$ as has been discussed above.

Once more we would like to note that in the present research the name
"effective charge" in this high voltage nonlinear regime is used just by
analogy with the low voltage regime and the precise meaning is given by the
ratio in Eq. (\ref{Eff_Ch}). However, as mentioned above, this ratio is
experimentally relevant and can be measured with high precision at any
voltage.

\section{Conclusion}\label{Conclusion}
In conclusion, we have explored strongly nonequilibrium Majorana fluctuations
of the electric current. It has been shown that in general these fluctuations
are characterized by two fractional effective charges $e^*_l$ and $e^*_h$ at
low and high energies, respectively. We have demonstrated that the low energy
effective charge $e^*_l$ might be washed out by thermal noise but the high
energy effective charge $e^*_h$ is robust and persists up to very high
temperatures. The latter, thus, represents a challenge for modern experiments
on noise phenomena in quantum dots since it is protected by high bias voltage
$V$ from all the perturbations whose strengths are smaller than $e|V|$. In
particular, electron-electron interactions and disorder will not change the
high energy effective charge if their characteristic energy scales, $V_{e-e}$,
$V_{dis}$, are smaller than $e|V|$, that is if $V_{e-e}<e|V|$ and
$V_{dis}<e|V|$. Of course, in future our research should be improved with more
realistic models to test the robustness of the high energy effective charge
and to predict its value when, {\it e.g.}, the density of states in the
contacts is not constant or multiple levels in the quantum dot are involved in
the transport. However, the model we have explored in the present research is
already quite standard and is often applied in many other contexts to
successfully describe modern experiments. We, therefore, believe that our
results, in particular, the high energy effective charge may become a reliable
platform for a unique signature of Majorana fermions out of strongly
nonequilibrium fluctuations. 

\section*{Acknowledgments}
We thank M. Grifoni and W. Izumida for important discussions. Support from the
DFG under the program SFB 689 is acknowledged.
\appendix
\section{Majorana field operators in the Keldysh partition function}\label{appA}
The partition function on the Keldysh closed time contour is usually
\cite{Altland_2010} constructed by splitting this contour into small time
intervals via writing the evolution operator as the product of elementary
evolutions between neighboring discrete times. One then inserts the coherent
state representation of the identity operator between all those elementary
evolution operators and calculates the matrix elements of the elementary
evolution operators between the fermionic coherent states $|\chi_{i-1}\rangle$
and $|\chi_i\rangle$ at neighboring discrete times $(i-1)$ and $i$. The
corresponding matrix elements of fermionic creation and annihilation
operators, $a^\dagger_\alpha$ and $a_\alpha$ ($\alpha$ is a single particle
index), may be written as \cite{Altland_2010}:
\begin{equation}
\langle\chi_i|a^\dagger_\alpha|\chi_{i-1}\rangle=\bar{\chi}_{i,\alpha}\langle\chi_i|\chi_{i-1}\rangle,\quad
\langle\chi_i|a_\alpha|\chi_{i-1}\rangle=\frac{\partial\langle\chi_i|\chi_{i-1}\rangle}{\partial\bar{\chi}_{i,\alpha}},
\label{matr_elem_cr_an}
\end{equation}
where $\chi_{i,\alpha}$ are the generators of the Grassmann algebra at
discrete time $i$.

Therefore, the Majorana operator fields $\zeta_{1,2}$ obtained using
Eq. (\ref{matr_elem_cr_an}) for the linear combinations of the Dirac fermion
operators, $(f^\dagger+f)$ and $i(f-f^\dagger)$, respectively, have the
following form at a given discrete time $i$ of the Keldysh closed time
contour:
\begin{equation}
\zeta_{1,i}=\bar{\chi}_i+\frac{\partial}{\partial\bar{\chi}_i},\quad
\zeta_{2,i}=i\biggl(\frac{\partial}{\partial\bar{\chi}_i}-\bar{\chi}_i\biggl).
\label{Maj_op_fld}
\end{equation}
Since the generators of the Grassmann algebra at a given discrete time $i$
satisfy the anticommutation relation
\begin{equation}
\biggl[\bar{\chi}_i,\frac{\partial}{\partial\bar{\chi}_i}\biggl]_+=1,
\end{equation}
the Majorana operator fields square to one at any given discrete time $i$ of
the Keldysh closed time contour,
\begin{equation}
\zeta_{1,i}^2=\zeta_{2,i}^2=1.
\label{Maj_op_fld_sqr_1}
\end{equation}
In terms of the Dirac operator fields $\bar{\chi}$, $\chi$, the Hamiltonian
$\mathcal{H}_{\rm TS}(q,p)$ from the main text takes the form at a given
discrete time $i$:
\begin{equation}
\mathcal{H}_{\rm TS}(q,p)=\xi\biggl(\bar{\chi}_i\frac{\partial}{\partial\bar{\chi}_i}-\frac{1}{2}\biggl).
\label{H_TS_Dir}
\end{equation}
The constant term in Eq. (\ref{H_TS_Dir}) cancels out on the forward, "$+$",
and backward, "$-$", branches of the Keldysh closed time contour and,
therefore, plays no role.
\section{Keldysh action}\label{appB}
Since the overlap of any two fermionic coherent states, $|\psi\rangle$ and
$|\phi\rangle$ has the form \cite{Altland_2010}:
\begin{equation}
\langle\psi|\phi\rangle=\exp\biggl(\sum_\alpha\bar{\psi}_\alpha\phi_\alpha\biggl),
\label{Ferm_coh_st_ovr}
\end{equation}
one can see from Eq. (\ref{matr_elem_cr_an}) that in the calculation of the
matrix elements of the elementary evolution operators the derivatives in
Eq. (\ref{Maj_op_fld}) bring the generators $\chi_{i-1}$ at the discrete times
neighboring to the discrete times $i$, {\it i.e}, from the Grassmann algebras
at the discrete times $(i-1)$.

Therefore, in the calculation of the matrix elements the Majorana operator
fields, $\zeta_{1,i}$ and $\zeta_{2,i}$, bring, respectively, in the continuum
limit the factors $(\bar{\chi}(t)+\chi(t))$ and $i(\chi(t)-\bar{\chi}(t))$,
$t\in\mathcal{C}_\mathcal{K}$, while the Hamiltonian in Eq. (\ref{H_TS_Dir})
brings the factor $\xi\bar{\chi}(t)\chi(t)$, where the constant term in
Eq. (\ref{H_TS_Dir}) is dropped out as explained above.

As a result, the Keldysh action $\mathcal{S}_\mathcal{K}$ from the main text
may be written as:
\begin{equation}
\mathcal{S}_\mathcal{K}=\mathcal{S}_0+\mathcal{S}_{\rm T},
\label{Keld_act}
\end{equation}
where $\mathcal{S}_0$ is the conventional noninteracting (quadratic) action of
the isolated quantum dot, contacts and topological superconductor and
$\mathcal{S}_{\rm T}$ is the action which describes the tunneling interaction
between the quantum dot and contacts as well as between the quantum dot and
topological superconductor. It has the following form:
\begin{equation}
\eqalign{\mathcal{S}_{\rm T}=
-\int_{\mathcal{C}_\mathcal{K}}dt\{\eta^*[\bar{\psi}(t)\chi(t)+\bar{\psi}(t)\bar{\chi}(t)]+{\rm H.c.}\}-\cr
-\int_{\mathcal{C}_\mathcal{K}}dt\sum_{l=\{L,R\};k_l}[T_l\bar{\phi}_{lk_l}(t)\psi(t)+{\rm H.c.}],}
\label{Tunn_act}
\end{equation}
where the notations are taken from the main text.

As one can see from Eq. (\ref{Tunn_act}), the only difference from the case
of a field integral without the Majorana operator fields,
Eqs. (\ref{Maj_op_fld})-(\ref{Maj_op_fld_sqr_1}), is the presence of anomalous
terms in Eq. (\ref{Tunn_act}) such as
$\bar{\psi}(t)\bar{\chi}(t)$. One deals with these terms in the same way as in
the field integral theory of superconductivity \cite{Altland_2010} where the
particle-hole space is introduced via the Nambu spinors. The additional
particle-hole index, however, is technically inessential because the whole
action is still quadratic and, therefore, is exactly solvable.
\section*{References}

\begin{thebibliography}{10}
\expandafter\ifx\csname url\endcsname\relax
  \def\url#1{{\tt #1}}\fi
\expandafter\ifx\csname urlprefix\endcsname\relax\def\urlprefix{URL }\fi
\providecommand{\eprint}[2][]{\url{#2}}

\bibitem{Brown_1828}
Brown R 1828 {\em New Phil. J., Edinburgh\/} {\bf 5} 358

\bibitem{Einstein_1905}
Einstein A 1905 {\em Ann. Phys., Paris\/} {\bf 17} 549

\bibitem{Smoluchowski_1906}
Smoluchowski M 1906 {\em Ann. Phys., Paris\/} {\bf 21} 772

\bibitem{Svedberg_1906}
Svedberg T 1906 {\em Z. Elektrochem.\/} {\bf 12} 853

\bibitem{Perrin_1908}
Perrin J 1908 {\em C.R. Acad. Sci., Paris\/} {\bf 146} 967

\bibitem{Nyquist_1928}
Nyquist H 1928 {\em Phys.\ Rev.\/} {\bf 32} 110

\bibitem{Callen_1951}
Callen H~B and Welton T~A 1951 {\em Phys.\ Rev.\/} {\bf 83} 34

\bibitem{Landau_V}
Landau L~D and Lifshitz E~M 1980 {\em Statistical Physics. Part 1: Course of
  Theoretical Physics\/} vol~5 (Pergamon Press)

\bibitem{Majorana_1937}
Majorana E 1937 {\em Nuovo Cimento\/} {\bf 14} 171

\bibitem{Fu_2008}
Fu L and Kane C~L 2008 {\em Phys.\ Rev.\ Lett.\/} {\bf 100} 096407

\bibitem{Fu_2009}
Fu L and Kane C~L 2009 {\em Phys.\ Rev.\ B\/} {\bf 79} 161408(R)

\bibitem{Lutchyn_2010}
Lutchyn R~M, Sau J~D and \text{Das Sarma} S 2010 {\em Phys.\ Rev.\ Lett.\/}
  {\bf 105} 077001

\bibitem{Oreg_2010}
Oreg Y, Refael G and von Oppen F 2010 {\em Phys.\ Rev.\ Lett.\/} {\bf 105}
  177002

\bibitem{Kitaev_2001}
\text{Yu} Kitaev A 2001 {\em Phys.-Usp.\/} {\bf 44} 131

\bibitem{Alicea_2012}
Alicea J 2012 {\em Rep. Prog. Phys.\/} {\bf 75} 076501

\bibitem{Flensberg_2012}
Flensberg K 2012 {\em Semicond. Sci. Technol.\/} {\bf 27} 124003

\bibitem{Sato_2016}
Sato M and Fujimoto S 2016 {\em J. Phys. Soc. Japan\/} {\bf 85} 072001

\bibitem{Albrecht_2016}
Albrecht S~M, Higginbotham A~P, Madsen M, Kuemmeth F, Jespersen T~S,
  Nyg{\r{a}}rd J, Krogstrup P and Marcus C~M 2016 {\em Nature\/} {\bf 531} 206

\bibitem{Mourik_2012}
Mourik V, Zuo K, Frolov S~M, Plissard S~R, Bakkers E~P~A~M and Kouwenhoven L~P
  2012 {\em Science\/} {\bf 336} 1003

\bibitem{Hewson_1997}
Hewson A~C 1997 {\em The Kondo Problem to Heavy Fermions\/} (Cambridge
  University Press, Cambridge)

\bibitem{Fidkowski_2012}
Fidkowski L, Alicea J, Lindner N~H, Lutchyn R~M and Fisher M~P~A 2012 {\em
  Phys.\ Rev.\ B\/} {\bf 85} 245121

\bibitem{Kundu_2013}
Kundu A and Seradjeh B 2013 {\em Phys.\ Rev.\ Lett.\/} {\bf 111} 136402

\bibitem{Ilan_2014}
Ilan R, Bardarson J~H, Sim H~S and Moore J~E 2014 {\em New J. Phys.\/} {\bf 16}
  053007

\bibitem{Lobos_2015}
Lobos A~M and \text{Das Sarma} S 2015 {\em New J. Phys.\/} {\bf 17} 065010

\bibitem{Wang_2016}
Wang R, Lu H~Y, Wang B and Ting C~S 2016 {\em Phys.\ Rev.\ B\/} {\bf 94} 125146

\bibitem{Lutchyn_2017}
Lutchyn R~M and Glazman L~I 2017 {\em arXiv:1701.00184\/}

\bibitem{Smirnov_2015}
Smirnov S 2015 {\em Phys.\ Rev.\ B\/} {\bf 92} 195312

\bibitem{Liu_2015}
Liu D~E, Cheng M and Lutchyn R~M 2015 {\em Phys.\ Rev.\ B\/} {\bf 91} 081405(R)

\bibitem{Liu_2015a}
Liu D~E, Levchenko A and Lutchyn R~M 2015 {\em Phys.\ Rev.\ B\/} {\bf 92}
  205422

\bibitem{Beenakker_2015}
Beenakker C~W~J 2015 {\em Rev.\ Mod.\ Phys.\/} {\bf 87} 1037

\bibitem{Valentini_2016}
Valentini S, Governale M, Fazio R and Taddei F 2016 {\em Physica E\/} {\bf 75}
  15--21

\bibitem{Sela_2006}
Sela E, Oreg Y, von Oppen F and Koch J 2006 {\em Phys.\ Rev.\ Lett.\/} {\bf 97}
  086601

\bibitem{Ferrier_2016}
Ferrier M, Arakawa T, Hata T, Fujiwara R, Delagrange R, Weil R, Deblock R,
  Sakano R, Oguri A and Kobayashi K 2016 {\em Nature Physics\/} {\bf 12} 230

\bibitem{Smirnov_2013}
Smirnov S and Grifoni M 2013 {\em New J. Phys.\/} {\bf 15} 073047

\bibitem{Smirnov_2013a}
Smirnov S and Grifoni M 2013 {\em Phys.\ Rev.\ B\/} {\bf 87} 121302(R)

\bibitem{Altland_2010}
Altland A and Simons B 2010 {\em Condensed Matter Field Theory\/} 2nd ed
  (Cambridge University Press, Cambridge)

\bibitem{Wang_2013}
Wang E, Ding H, Fedorov A~V, Yao W, Li Z, Lv Y~F, Zhao K, Zhang L~G, Xu Z,
  Schneeloch J, Zhong R, Ji S~H, Wang L, He K, Ma X, Gu G, Yao H, Xue Q~K, Chen
  X and Zhou S 2013 {\em Nature Physics\/} {\bf 9} 621

\bibitem{Vernek_2014}
Vernek E, Penteado P~H, Seridonio A~C and Egues J~C 2014 {\em Phys.\ Rev.\ B\/}
  {\bf 89} 165314

\bibitem{Goldhaber-Gordon_1998}
Goldhaber-Gordon D, Shtrikman H, Mahalu D, Abusch-Magder D, Meirav U and
  Kastner M~A 1998 {\em Nature\/} {\bf 391} 156

\bibitem{Haim_2015}
Haim A, Berg E, von Oppen F and Oreg Y 2015 {\em Phys.\ Rev.\ B\/} {\bf 92}
  245112

\end{thebibliography}

\providecommand{\newblock}{}

\end{document}